\documentstyle[11pt]{article}
 
 \newcommand{\degr}{^\circ}
 
\title{\vspace{4cm}\large\bf THE HUBBLE FLOW: \\
WHY DOES THE COSMOLOGICAL EXPANSION PRESERVE\\ 
ITS KINEMATICAL IDENTITY FROM A FEW MPC DISTANCE\\
TO THE OBSERVATION HORIZON?
}
\author{Igor D.~Karachentsev$^1$,
Arthur D.~Chernin$^{2,3,4}$, Pekka Teerikorpi$^2$\\
$^1$Special Astrophysical Observatory, Nizhnii Arkhys, 
369167, Russia,\\
$^2$Tuorla Observatory, Turku University, Piikki\"o, 
21 500, Finland,\\
$^3$Sternberg Astronomical Institute, Moscow 
University, Moscow, 119899,
Russia,\\
$^4$Astronomy Division, University of Oulu, 90014, 
Finland
}

\date{~}

\begin{document}

\maketitle

\begin{abstract}
\noindent
The problem of the physical nature of the Hubble 
flow in the Local Volume ($D < 10$ Mpc) stated 
by Sandage(1986, 1999) is studied. New observational 
data on galaxy motions and matter distribution around 
the Local Group and nearby similar systems are 
described. 
Dynamical models are discussed on the 
basis of the recent data on cosmic vacuum or dark energy.
\end{abstract}
\vfill
\eject

\section{Introduction}

In a recent paper, Sandage (1999) has
emphasized evidence that the rate of the cosmological
expansion in the Local Volume ($D < 10$ Mpc) is
`similar, if not precisely identical', to the global 
rate (see also Sandage 1986, Teerikorpi 1997, 
Ekholm et al. 1999, Giovanelli et al. 1999). 
This is a severe challenge to the current cosmological 
theories, especially  in view the fact that the linear  
expansion flow 
starts from the distances of a few Mpc from the 
Local Group (Sandage et al. 1972, Karachentsev et al. 2001, 2002,
Ekholm et al. 2001).  
Indeed, why is the galaxy velocity field fairly 
regular in the area where the galaxy spatial 
distribution is very irregular? And how can it 
be compatible with the bulk motion of the volume 
with a high velocity of about 600 km$s^{-1}$?

In this paper, we 
first
describe
new observational data on the kinematics and distribution of galaxies  
in the Local Volume and
then suggest a theoretical framework 
which appears to offer a possible solution to the
above mentioned problems. 
This approach is suggested by the recent discovery of the  
cosmic vacuum, or dark energy, and on the data on its 
energy density (Riess et al. 1998, Perlmutter et al. 1999).

We note that the Local Volume is in many ways optimal for study of
questions concerning the various components of the universe and their
dynamics.  Here we have relatively
accurate distances and a good knowledge of the distribution
of galaxies which turns out to be typical for the galaxy universe in general.
In this volume the Hubble law starts and one may
see both its linear form and to measure its dispersion. One may also
detect minor deviations due to the differential
velocity field caused by the Virgo cluster.  The Local Volume is also
deep inside the (unknown) volume which has the zero velocity
relative to the cosmic background radiation.  With its very clumpy
galaxy distribution  it is also deep inside
the volume in which the distribution may be regarded as uniform.

\section{Global expansion rate and cosmic vacuum}

The global rate of the expansion is the rate on the spatial scales 
of a hundred Mpc and larger where the spatial matter 
distribution is considered smooth and uniform,   
on average (see, for instance, Wu, Lahav \& Rees 1999). 
This is the realm of the standard isotropic cosmology which views 
the linear expansion flow as a direct consequence of the 
uniformity of matter distribution. 

The precise value of the global expansion rate, or the 
Hubble constant, is still under discussion. However a 
value $H_0 = 70 \pm 6$ km s${-1}$/Mpc appears to cover 
most of the recent determinations on large ($> 200$ Mpc) and
"intermediate" (30-200 Mpc) scales (Theureau et al. 1997; Sandage 1999;
Giovanelli et al. 1999; Freedman et al. 2001). 

As to the theory, according to 
the Friedmann model $H = {\dot a}/a
$, where $a(t)$ is the scale factor of the model.   
For cold (non-relativistic) 
matter with the density $\rho_M$ and cosmological 
constant, or vacuum with the density $\rho_V$, the 
scale factor is given by the Friedmann 
equation:
\begin{equation}
\dot a^2 = {\frac{8\pi}{3}} G (\rho_M + \rho_V) a^2 - k.
\end{equation}

For the spatially flat model ($k = 0$), the solution has a form:

\begin{equation}
a(t) = a_0 q^{-1/3} \sinh^{2/3}(\frac{3}{2} \alpha t),
\end{equation}
\noindent
where $a_0 = a(t_0)$ is the present-day value of $a(t)$,  
$q = \rho_V /\rho_M$, and $\alpha = (\frac{8\pi}{3} 
G \rho_V)^{1/2}$. Then 
\begin{equation}
H(t) = \alpha (\frac{3}{2} \alpha t).
\end{equation}

On the other hand, the Hubble constant is expressed from Eq.(1) 
directly in terms of $\alpha$ and $q$:
\begin{equation}
H = \alpha (\frac{1+q}{q})^{1/2}.
\end{equation}

For the present-day universe, the `concordant'
observational figures for the densities are:
\begin{equation}
\Omega_V = \rho_V/\rho_c = 0.7 \pm 0.1; \;\;\;
\Omega_M = \rho_M/\rho_c = 0.3 \pm 0.1, 
\end{equation},
where $\rho_c$ is the critical density. 
As a result, one has a rather narrow interval for the 
present-day global rate in the flat ($\Omega_M + \Omega_V = 1$) 
model: 
\begin{equation}
 H_0 \le (1.2 \pm 0.4) \alpha.  
\end{equation}

Actually, an approximate `concordant' estimate 
is possible for all the three figures involved in 
the equations above (see for a review Chernin 2001):

\begin{equation}
 H_0 \sim t_0^{-1} \sim \alpha. 
\end{equation}

More precise numbers follow from the concordant observational 
evidence: 
\begin{equation}
H_0 = (2\pm 0.2) \times 10^{-18} s^{-1}; \;\; t_0 = 15 \pm 1 \;Gyr = (4.5 \pm
0.3) \times 10^{17} s; \;\; \alpha = (1.4 \pm 0.4) \times 10^{-18} s^{-1}.
\end{equation}

In real observations, the global 
expansion
rate appears as a {\em mean} 
value of many individual measurements. There are naturally 
some definite deviations from the mean value; but they are 
not too significant on the global scales. Indeed, galaxies, 
their groups, clusters and superclusters are able to produce 
peculiar velocities within the expansion flow in their vicinity; 
the absolute value of these velocities are practically within 
the interval 100-1000 km/s. Because of this, any real deviations 
are all below the 10\% level for the distances of 200 Mpc and 
larger.       

Thus we see that the concordant model enables one to 
put the observational data on the global expansion  
rate (see above) in general agreement with 
the cosmic age and the vacuum density. The model 
indicates that in the  
present state of the universe, the global rate is 
determined in terms of $\alpha$ by the vacuum alone 
with a practically perfect accuracy.   
This new conclusion  follows directly from the discovery 
of the cosmic vacuum with its energy density which is larger than 
the total energy density of all the other forms of cosmic matter. 

\section{Matter distribution in the Local Volume}

Galaxies are distributed on the sky very inhomogeneously. This basic
property of galaxies had been known long before their extragalactic nature
was established. Clustering of galaxies towards each other is seen in a
wide range of scales: from a typical galaxy diameter, $\sim$10 kpc, up to a
scale of $\sim$30 Mpc exceeding a supercluster dimension. New optical and
infrared sky surveys led to the conclusion that galaxy distribution is not
homogeneous even on a scale of $\sim$500 Mpc (Busswell et al. 2003), which
reaches about 1/10 of the horizon of the universe. Over the last decades,
old nomenclature has described small and large galaxy systems: pairs,
groups, clusters and superclusters, was updated with an idea of the large
scale structure consisting of cosmic ``filaments" and "walls'' framing
giant empty volumes.

Generally, in the Local Volume one may see examples of all the
components visible on larger scale 3-D maps: groups, elongated structures,
filaments and voids. In addition, the local ($< 10$ Mpc) spatial distribution
appears to be fractal, with $D \approx 1.8$ (Tikhonov 2002), in agreement
 with the distribution on larger scales, up to about 100 Mpc
(Sylos-Labini et al. 1998; Teerikorpi et al. 1998; Wu, Lahav \& Rees 1999;
Baryshev \& Teerikorpi 2002).

  All mentioned properties of the large scale structure are seen in
Fig. 1, which presents the sky distribution of 5272 nearest galaxies
in the equatorial coordinates. They are selected from the last version of
the Lyon Extragalactic Database (=LEDA) (Paturel et al. 1996) by the
condition that their corrected radial velocities is $V_{LG} <$ 2300 km/s.
The galaxies are shown as filled circles with sizes inversly
proportional to their distances (radial velocities). The gray belt
corresponds to the Zone of Avoidance in the Milky Way (galactic latitude
of $\pm 10\degr$), where the lack of galaxies is caused by strong Galactic
extinction. As seen in the figure, the nearby galaxies are concentrated
towards the Local Supercluster plane, and the Virgo cluster is the most
dense part of it. The Virgo cluster is located near the center of Fig. 1
(marked with the character ``V'') and has a distance of 17 Mpc from us. The distribution
of these galaxies in the supergalactic coordinates is presented in Fig. 2.
About  half of the galaxies within the radius of 32 Mpc is situated
in the Local Supercluster disk.

In the southern supergalactic hemisphere
(SGL $\sim250\degr$, SGB $\sim-40\degr$) another less rich cluster of galaxies,
Fornax, alined along the supergalactic longitude is seen. The Fornax cluster
has a distance of 20 Mpc. In the northern hemisphere there is a significant
deficit of galaxies with radial velocities $V_{LG} <$ 1500 km/s. This almost
empty volume in Hercules- Aquila with a linear diameter of 20 Mpc was called
``Local Void'' by Tully (1988). In the opposite direction (the Orion
constellation) there is another smaller empty region called ``Local Minivoid''
(Karachentsev et al. 2002).

    Projection of nearby and distant galaxies onto the sky makes difficult
viewing the 3D structure of the Local volume. Passing to the Cartesian
supergalactic coordinates allows us to see the local relief in a new aspect.
Fig. 3 presents the distribution of galaxies with radial velocities $V_{LG}
<$ 1500 km/s in projection onto the supergalactic plane. The radial
velocity of each galaxy, $V_{LG} = H_0\times D$, is used as the galaxy distance, D,
where the Hubble parameter  $H_0$ = 72 km s$^{-1}$Mpc$^{-1}$ is assumed from modern data.
The central part of Fig. 3 shows the distribution of nearby galaxies
situated within the Local supercluster plane. The thickness of this slice
is taken to be $\pm$300 km/s along the SGZ axis. The upper (a) and the bottom
(c) pannels of Fig.3 present the distribution of remaining galaxies situated
above (SGZ $> $300 km/s) and below (SGZ $< -$300 km/s) the Supercluster disk,
respectively. The main feature of the local landscape, the Virgo cluster,
is elongated approximately along the +SGY axis. To a considerable extent,
its alongation is fictituous, being caused by internal virial motions with
a dispersion  $\sigma_v$ = 650 km/s. Apart from the Virgo cluster, in the
Supergalactic plane there are other more scattered structures: the Ursa
Majoris cloud, the Canes Venaticy cloud, the Triangulum spur, etc., which
have been revealed by Tully owing to their contrast above the average
number density. The map of the southern supergalactic hemisphere shows
the disposition of nearby galaxy clouds in Fornax, Leo, and Antlia. The
presence of the Local Void is well seen in the galaxy distribution North
of the Supergalactic disk.

  It appears that our Galaxy is located not in the richest, nor in the
poorest region of the Local supercluster. Kraan-Korteweg \& Tammann (1979)
proposed to call the space region around the Local Group with radial
velocities of galaxies $V_{LG} <$ 500 km/s ``Local Volume''. After ommiting
the Virgo cluster members having $V_{LG} <$ 500 km/s because of their virial
motions, the Local Volume population contains 179 galaxies, being
rather representative in number. During the last years, special effort has
been undertaken to increase the Local Volume population. ``Blind'' surveys
of the sky in the 21 cm line (Ryan-Weber et al. 2002), infrared and radio
surveys of the Zone of Avoidance (Kraan-Korteweg \& Lahav, 2000) and searches
for new dwarf galaxies of very low surface brightness based on the POSS-II
and ESO/SERC plates (Karachentseva \& Karachentsev, 1998, 2000) led to the
increasing of the total number of the Local Volume galaxies more than
two times.

  Radial velocities of galaxies, especially situated within groups and
clusters, give only an approximate estimate of distances to these galaxies.
That was a reason to initiate a vast program of distance measuring to
nearby galaxies, independently from their radial velocities. Over the
last 10--15 years many nearby galaxies have been resolved into stars
for the first time. The luminosity of their brightest blue and red stars
have been used to determine galaxy distances with a typical accuracy of
(20 -- 25)\% (Karachentsev \& Tikhonov, 1994; Karachentsev et al., 1997).
Later, the distance measurement error was decreased to $\sim$10\% by
using the luminosity of the tip of the red giant stars. This
labour- consuming program requiring a lot of observing time with the
largest ground-based telescopes, as well as the Hubble Space Telescope,
is yet not complete. So far, the distances have been measured for about 150
galaxies situated within 6 Mpc (Karachentsev et al. 2003). The distribution
of these galaxies is presented in Fig.4. The Local Volume galaxies are
projected onto the Supergalactic plane, SGX, SGY, and shown as filled
circles. In this region there are eight known groups whose principal
galaxies: the Milky Way, M 31, IC 342, M 81, Cen A, M83, NGC 253, and M 94
are indicated with asterisks. Comparing the true 3D map of the Local
Volume in Fig.4 with its approximate analogy in the redshift space (Fig.3),
we recognize a higher density contrast of groups in Fig.4 and also
the absence of the ``virial tail'' directed towards the Virgo cluster.

    Note that the total number of 5272 galaxies, being averaged over the
D = 32 Mpc volume, yield the expected number of galaxies  within
D = 6 Mpc to be 35. This number is 7 times as low as their observed
number in the Local Volume. However, the excess is caused completly by
the faintest galaxies unseen in more distant regions of the Local
Supercluster.

\section{Galaxy kinematics in the Local Volume.}

Extensive measurements of distances to galaxies independent of their radial
velocities provide us with a possibility to study the peculiar velocity field
on different scales. Analysing the peculiar velocity map, we can establish
reasons (inhomogeneities of gravitational potential) which generate the
observed deviations in galaxy motions with respect to the regular Hubble
flow. Surprisingly, such a kind of data on very nearby galaxies have turned
out to be known over the last 2--3 years only!

  A sample of observational data on radial velocities and distances for
$\sim$150 nearby galaxies given by Karachentsev et al. (2003) is presented
in Fig. 5. The galaxies with accurate ($\sim$10\%) distances measured from the
luminosity of cepheids or red giants are shown by filled circles. The
galaxies with less reliable distances (via the brightest stars or
Tully-Fisher relation) are indicated by crosses. The radial velocities
of galaxies are reduced to the Local Group centroid. The solid line in
Fig. 5 corresponds to the Hubble relation, $V_{LG}= H_0\times D$ with $H_0 =
72$ km s$^{-1}$Mpc$^{-1}$, when a decelerating gravitational action of the
Local Group mass, $1.3\cdot 10^{12} M_\odot$, is taken into account. Apart
from the presented galaxies, the Local Volume of radius $D <$ 6 Mpc contains
about 100 galaxies, whose distances are still unknown.

  The largest deviations from the Hubble relation take place for the
galaxies situated within two nearby groups around M 81 and Cen A.
The members of these groups are shown in Fig.5 by open circles and open
squares, respectively. As it was noted by Karachentsev \& Makarov (1996,
2001), significant deviations from the regular Hubble flow are caused
by anisotropic expansion of the Local Volume. The local value of the Hubble
parameter can be described by a tensor  $H_{ij}$ with the main axial ratios
(81$\pm$3) : (62$\pm$3) : (48$\pm$5) km s$^{-1}$Mpc$^{-1}$. Here the minor axis of the Hubble
ellipsoid is directed towards the Local Supercluster poles, and the major
axis is pointed (29$\pm$5) deg away from the direction to the Virgo cluster.
The nature of this phenomena remains unclear yet. In any case,
it does not agree with the idea of spherically- symmetric Virgo-centric
flow, which has been discussed by many authors.

  The most enigmatic property of the local Hubble flow turns out to be its
``coldness''. According to Sandage et al. (1972), the typical random velocity
of galaxies is 70 km/s. Just the same value has been derived by
Karachentsev \& Makarov (1996) for galaxies within 7 Mpc from us. Later,
Karachentsev \& Makarov (2001) showed that for the nearest galaxies
with $D < 3$ Mpc their radial velocity dispersion does not exceed 30 km/s.
Also, using Cepheid distances only, Ekholm et al. (2001) derived a local
dispersion of 40 km/s. These results are as was predicted by Sandage:
the smaller the distance measurement
error the lower the observed peculiar velocity dispersion.

  In the Local Group and in the other nearby groups the characteristic
virial velocity is also about 70 km/s. However, the group centroids
themselves have much lower chaotic motions. Fig. 6 presents the Hubble
diagram for centroids of the eight nearby groups shown in Fig. 4.
Their velocities and distances are taken with respect to the Local Group
centroid situated between M 31 (Andromeda) and our Galaxy. It appears
that the group centroids have a scatter of only 29 km/s in regard to the
Hubble relation with $H_0$ = 72 km s$^{-1}$Mpc$^{-1}$. The distance measurement errors
are shown by horizontal bars. As seen, for the galaxy groups with radial
velocities $V_{LG} \sim$250 km/s their $\sim$10\% distance errors lead to the Hubble
velocity errors $\sim$25 km/s comparable with the observed value of $\sigma_v$.

  Finally, the most complete and accurate data on radial velocities and
distances of nearby galaxies demonstrate that the local Hubble flow has
almost the same value of the Hubble parameter as the global flow: $H_0 =
(71\pm4)$ km s$^{-1}$Mpc$^{-1}$ (Spergel et al. 2003).
However, some uncertainty in the classically measured Hubble constant
remains until the suspected extragalactic Cepheid distance bias
is fully investigated (Teerikorpi \& Paturel 2002).

\section{ Where are we moving towards?}

Together with our Sun and our Galaxy we take part in different cosmic
motions whose value and direction have been discussed by many authors.
The initial data on these motions were controversal because of the low
quality of determination of galaxy distances. However, at the present time,
one can recognize a rather concordant picture of cosmic motions described
below.

  Taking part in the rotation of the Galaxy, our Sun moves at a velocity
of (220$\pm20)$ km/s towards $l = 90\degr$, $b = 0\degr$ in the galactic
coordinates (Vaucouleurs et al. 1991). Apart from the regular circular rotation
the Sun has its individual velocity of 16 km/s towards $l = 53\degr$,
$b = 25\degr$  with respect to surrounding stars
( Vaucouleurs et al. 1991). Considering
velocities and distances of nearby galaxies, Karachentsev \& Makarov
(1996, 2001) established that the Sun moves with respect to the Local
Group centroid with the velocity (316$\pm$11) km/s in the direction
$l = (93\pm2)\degr$, $b = (-4\pm1)\degr$. When these vectors are substracted,
we derive that the motion of the Galaxy center with respect to the Local Group
centroid is 91 km/s towards $l = 163\degr$, $b = -19\degr$. As it is known,
the centers of the Galaxy and Andromeda (M 31) are approaching each
other at a velocity of $-$120 km/s. If the Galaxy mass is twice as low as
the M 31 mass, our Galaxy should move towards M 31 at a velocity of
80 km/s. After excluding this expected velocity component, a residual
(random) velocity of the Galaxy is only 23 km/s towards $l = 56\degr$,
$b = 0\degr$. Its value and direction can be easely explained by a not
strickly radial motion of the Galaxy towards M 31 or by an underestimated
circular velocity of the Galaxy in the Sun's neighbourhood.

  Measurements of the dipole anizotropy of the cosmic microwave
background (CMB) showed that our Sun moves with respect to the CMB
at a velocity of (370$\pm$3) km/s towards $l = (264.4\pm0.3)\degr$, $b =
(48.4\pm0.5)\degr$ (Kogut et al. 1993). Therefore, in the absolute frame
(CMB) the Sun's motion is known with an accuracy of better than 1\%.
Because of the Sun's motion with respect to the Local Group and its motion
with respect to the CMB have nearly opposite directions, the
velocity of the Local Group centroid itself with respect to the CMB
has a huge value, (634$\pm$12) km/s, towards $l = (269\pm3)\degr$, $b =
(48.4\pm0.5)\degr$. The origion of such a fast motion of the Local Group
was a puzzle for many observers trying to determine the Local Group
velocity with respect to nearby and distant galaxies. The most defined
results were obtained by Tonry et al. (2000) who measured accurate
($\pm$10\%) distances to 300 early-type galaxies with radial velocities
$<$ 3000 km/s. An analysis made from these observational data reveals
that the Local Group takes part in different kinds of motion:

  a) towards the Virgo cluster center ( $l = 274\degr$, $b = 75\degr$) at
a velocity of 139 km/s,

  b) towards so-called ``the Great Attractor'' in Hydra-Centaurus
( $l = 291\degr$, $b = 17\degr$, $D$ = 44 Mpc) at a velocity of 289 km/s, and

  c) in the direction away from the Local Void (i.e. towards $l = 228\degr$,
$b = -10\degr$) at a velocity of $\sim$200 km/s.

  When all the three motions have been taken into account,
the residual velocity of
the Local Group towards $l = 281\degr$, $b = 43\degr$ is only 166 km/s.
According to Tonry et al. (2000), the error in the residual velocity is about
120 km/s that is why they considered the Local Group to be practically at rest
relative to remote galaxies.

  The bulk galaxy motion within a radius of $\sim(100--200)$ Mpc with respect
to the CMB was studied by different observational teams. Most of the
approaches relied on the Tully-Fisher relation in estimating
distances to spiral galaxies from the amplitude of their internal motions.
As a result, Giovanelli et al. (1998) and Dekel et al. (1999) derived
the bulk motion parameters: $V $= 200 km/s, $l = 295\degr$, $b = 25\degr$, and
$V$ = 370 km/s, $l = 305\degr$, $b = 14\degr$, respectively. To study coherent
large-scale motions, Karachentsev et al. (1993) created a special catalog
of flat disk-like galaxies seen edge-on. This catalog (FGC) covers
homogeneously the whole northern and southern sky. Based on the FGC sample,
Karachentsev et al. (2000) derived the dipole solution: $V = (300\pm75$) km/s,
$l = (328\pm15)\degr$, and $b = (7\pm15)\degr$. Applying for the FGC sample of
new photometric data from the 2MASS sky survey yields the following
parameters of the galaxy bulk motion: $V = (199\pm61$) km/s, $l = (301\pm18)\degr$,
and $b = (-2\pm15)\degr$ (Kudrya et al. 2003).

\begin{table}
\topmargin=-4cm
\caption{Cosmic motions of the Sun, the Galaxy, and the Local Group}
\begin{tabular}{lrrrrrrl}\hline

\multicolumn{1}{l}{Motion type}&
\multicolumn{1}{c}{V}&
\multicolumn{1}{c}{l}&
\multicolumn{1}{c}{b}&
\multicolumn{1}{c}{Vx}&
\multicolumn{1}{c}{Vy}&
\multicolumn{1}{c}{Vz}&
\multicolumn{1}{c}{Note}\\
&\multicolumn{1}{c}{km/s}&
\multicolumn{2}{c}{$\degr$}&
\multicolumn{3}{c}{ km/s}& \\ \hline
Sun vs. LSR            &    16  &   53& +25  &   9&  12  &   7 & Vaucouleurs et al.(1991)   \\
Galactic rotation      &   220  &   90&   0  &   0&  220 &   0 & Vaucouleurs et al.   \\
		       & $\pm20$ &     &      &    &      &     & (1991), $R_\odot$ = 8 kpc\\
Sun vs. LG centroid    &   316  &   93&  $-$4  & $-$16&  315 & $-$22 & Karachentsev, Makarov\\
		       & $\pm11$  &$\pm2$& $\pm1$  &    &      &     &    (1996, 2001)      \\
MW vs. LG              &    91  &  163& $-$19  & $-$25&   83 & $-$29 &                      \\
MW vs. M31             &    80  &  121& $-$23  & $-$38&   64 & $-$29 & expected             \\
residual MW            &    23  &   56&   0  &  13&   19 &   0 & non-radial orbit ?   \\
			 \hline
Sun vs. CMB            &   370  &  264&  48  & $-$24& $-$244 & 276 & Kogut et al. (1993)   \\
		       & $\pm3$  & $\pm.3$ & $\pm.5$ &    &      &     &                       \\
			 \hline
LG vs. CMB             &   634  &  269&  28  &  -8& $-$559 & 298 &                       \\
		       & $\pm12$  & $\pm3$ & $\pm1$ &    &      &     &                       \\
LG vs. Virgo           &   139  &  274&  75  &   3&  $-$36 & 134 & $D_{Vir}$=17 Mpc,Tonry et al.(2001) \\
LG vs. Great Attractor &   289  &  291&  17  &  98& $-$258 &  86 & $D_{GA}$=44 Mpc,Tonry et al.(2001) \\
LG vs. anti-Local Void &   200  &  228& $-$10  &$-$132& $-$146 & $-$36 & Local Void,$D\sim$20 Mpc \\
residual LG (-VA-GA+LV)&  166   & 281 & 43   & 23 &-119  &114  &                       \\
			 \hline
		       &        &  278&  38  &    &      &     & 2MASS gg centroid     \\
		       &        &  258&  30  &    &      &     & IRAS gg centroid      \\
		       &        &  315&  30  &    &      &     & Shapley concentration \\
			\hline
bulk vs. CMD           &   200  &  295&  25  &  77& $-$164 &  85 & Giovanelli et al.(1998),$D\sim$90 Mpc \\
bulk vs. CMD           &   370  &  305&  14  & 206& $-$294 &  90 & Dekel et al.(1998),$D\sim$70 Mpc   \\
bulk vs. CMD           &   300  &  328&   7  & 252& $-$157 &  37 & FGC, $D\sim$100 Mpc,        \\
		       & $\pm75$  & $\pm15$ & $\pm15$ &   &    &   & Karachentsev et al. (2000) \\
bulk vs. CMD           &   199  &  301&  $-$2  & 102& $-$170 &  $-$7 & FGC+2MASS, $D\sim$150 Mpc   \\
		       & $\pm61$  & $\pm18$ & $\pm15$  &    &      &     & Kudrya et al. (2003)  \\
			      \hline
\end{tabular}
\end{table}

  A survey of galaxy motions on different scales is presented in Table 1
and Fig. 7. Apart from M 31, Virgo (VA) and the Great Attractor (GA),
the position of the centroid of IRAS sources (Rowan-Robinson et al. 2000),
2MASS sources (Maller et al. 2003), and the Shapley concentration of rich
clusters (ShC) with its typical distance of $\sim$13000 km/s are shown
in the galactic coordinates. As seen from the map, all the large-scale
attractors (gray circles), as well as the apexes of bulk galaxy motions
(crosses) are concentrated in an approximately the same sky region,
cosmic ``Bermudas trianglum''. Such an association of apexes and attractors
can be easely understood if the distribution of Dark Matter on large scales
follows the galaxy distribution. It should be emphasized that the residual
velocity of the Local Group, $V $= 166 km/s, $l = 281\degr$, $b = 43\degr$,
(indicated in Fig. 7 as a diagonal cross) is directed almost towards the
clustering dipole of the 2MASS sources, $l = 278\degr$,$b = 38\degr$.
It means that the residual motion of the Local Group with respect to the CMB
can be generated by the large-scale structure seen in the 2MASS survey.

\section{Dynamic background in the Local Volume}

Turning to the theory,  we argue now that the local expansion 
rate could be due to the dynamical effect of the  
vacuum. 

>From the data on the matter distribution in the
Local Volume (Sec.3), one can see that the bulk
of mass (this is mostly dark matter) is concentrated in several groups like
the Local Group, if one consider the distances $1 \le R \le 7$. 
Matter dominates 
dynamically near the Local Group, while outside this 
region vacuum must dominate. A rough, but obvious and 
robust estimate shows this. 

Indeed, the mass of the Local Group $M_{LG}$  
is less than the effective gravitating mass of the vacuum 
in a surronding volume of   
 the size (radius) $R$, if $R$ is large enough:
\begin{equation}
M_{LG} < \frac{4\pi}{3} 2\rho_V R³,
\end{equation}
(here we take into account that the effective 
gravitating density of the vacuum is $-2 \rho_V$) 
and the vacuum dominates at distances 
\begin{equation}
R > R_V = (\frac{3 M_{LG}}{8\pi \rho_V})^{1/3}.
\end{equation}       
With $M_{LG} = (1.5 - 2.0)\times 10^{12} M_{\odot}$, 
and the vacuum density of Sec.2, one has (Chernin 2001
Baryshev et al. 2001):
\begin{equation}
R_V = 1.5 - 2.0 Mpc, 
\end{equation}
and the dynamical effect of vacuum dominates 
at distances of a few Mpc and larger.

A detailed computer model that takes into 
account the motion of the two major galaxies 
of the Local Group shows (Dolgachev et al. 2003)
that the surface of "zero gravity" where the 
gravity of the Local Group is exactly balanced by 
the anti-gravity of the cosmic vacuum is very near 
to sphere with the radius of 1.8 Mpc. This sphere  
change very slowly during the life time of the 
Group. It means that outside this surface the 
gravitational potential is spherically symmetrical 
and static (practically) for the last 12-13 Myr.   

It is significant that vacuum domination in the Local 
Volume is at least as strong as on the average, all  
over the Universe (i.e. on the global spatial scales). 
One may see this in terms of the effective mass that 
produces gravity at a given distance $R$ from the barycenter of 
the Local Group. Two kinds of estimates can be made for this. 
The first and simplest assumes that all the mass 
in the volume (up to a certain distance) is collected in 
the Local Group within a region of $\sim 1$ Mpc in size. 
If so, the ratio of the matter mass, $M_M = M_{LG}$, to 
the vacuum mass in the volume of the size (radius) $R$ 
around the Local Group scales with $R$ as       
\begin{equation}
M_V/M_M \propto (R/R_V)^3,   
\end{equation}
in accordance with Eq.(8). It means that at an intermediate distance of, say, 
$R = 3$ Mpc the mass ratio is $M_V/M_M \simeq 5$.

A more accurate estimate may take into account the  contribution
to the matter mass $M_M$ by the galaxies (and intergalactic matter)
distributed around the Local Group. According to the data of Sec.3,
the mass distribution is fractal, and  $M_M \propto R^{D}$,
where $0 < D \le 3$. In this case,
\begin{equation}
M_V/M_M \propto (R/R_V)^{D-3}.
\end{equation}
At a distance $R = 3$ Mpc (as above), the ratio is now
$M_V/M_M \simeq 1.8$, if $D = 2$, and
$M_V/M_M \simeq 3.1$, if $D = 1$.
For the most popular fractal dimension $D = 1.8$ (Peebles 1993,
Tikhonov et al. 2000) extended to, say, 20 Mpc,
the mass ratio $M_V/M_M$  is larger than the global effective ratio 14/3 in
almost all ($>$ 95\% !) the volume of space with this (20 Mpc) radius.
This means that dynamical dominance of the vacuum
in the Local Volume is actually even stronger than on the global scales.

\section{Expansion rate in the Local Volume}

The considerations above suggest that the present-day dynamics in the Local Volume 
outside the zero-gravity sphere can be 
considered as controlled by the vacuum alone, -- with the same (at least) accuracy 
as on the global scales. This enables one to study trajectories in the Local 
Volume neglecting the dynamical effect of matter, in the first (and main) 
approximation. In addition, the one may consider spherically symmetrical 
trajectories as a good approximation to the real motion of small galaxies 
in this volume.

In this approximation, the radial component of the equation of motion  
in the reference frame of the Local Group barycenter has a simple form:
\begin{equation}
{\ddot R} = \alpha ^2 R.
\end{equation}
 
The solution to the equation may be written as 
\begin{equation} 
R (t) = R_0(\chi) F(t, \chi),  
\end{equation}
where 
\begin{equation}
F = \exp [\alpha (t + T(\chi)], \;\;  \cosh [\alpha (t + T(\chi)], \;\; 
\sinh [\alpha (t + T(\chi)], 
\end{equation}
for parabolical, hyperbolical and elliptical trajectories, 
correspondingly.  

The solution describes the radial trajectory of 
a body (a dwarf galaxy) with the Euler radial coordinate $R$ 
and Lagrangian coordinate $\chi$. The solution  
is exact and nonlinear. The solution is also general in 
the sense that it contains two arbitrary functions of the Lagrangian 
coordinate, $R_0(\chi)$ and $T(\chi)$, and so it can fit all (reasonable)  
initial conditions for positions and velocities at the start of the motion.    

The solution describes regular  `unperturbed' Friedmann-Hubble trajectories,   
if $R_0(\chi) = \chi, T(\chi) = 0$.  In its general form, the solution 
describes a `perturbed' trajectory with arbitrary $R_0(\chi)$ and $T(\chi)$.  
The solution is valid (practically) since the time of the formation of the 
Local Group, i.e. since the moment $t_1 \simeq 1-3$ Gyr when the most of the 
material in the voulume was assembled into the two major galaxies of the 
Local Group.  

The solution gives the rate of expansion ${\dot R}/R$ measured for a given 
trajectory 
at a moment $t$ as a function of both $t$ and $\chi$. For a regular (unperturbed) 
trajectory the rate is simply $H_0 = \alpha$, as it is in the global solution (Sec.2).   
For a perturbed trajectory one may use, for instance, a hyperbolical solution: 
\begin{equation}
H (t, \chi) = {\dot R}/R = \alpha  [\alpha (t + T(\chi))].
\end{equation}
The dependence on $\chi$ is due to perturbations described by the 
arbitrary function $T(\chi)$; 
the other arbitrary function $R_0(\chi)$ does not enter this relation.  
The expansion rate does not depend on $\chi$  and coincides with the regular 
one, $H_0 = \alpha$,  
in the limit of large times; in this limit, the perturbations vanish.

On the contrary, at the moment of the Local Group formation, $t=t_1$, 
and soon after that, most of the 
trajectories might be  highly disturbed, so that for a typical trajectory 
the rate of expansion was significantly different from $\alpha$. 
And nevertheless big initial perturbaations are compatible with the 
present rather regular linear flow. For example, if $ T(\chi) = 0.2/\alpha$, 
then $H(t_1) \simeq 3 \alpha$ initially ($t_1 \simeq 0.1 1/\alpha$), while  
the present expansion rate $\simeq \alpha$ for the same $T$.  

Another simple solution can easily be obtained for radial trajectories in the case 
when gravity of matter is taken into account and the motion is parabolical.  
The solution has a form of Eq.(2), where one has now  $t + T(\chi)$ instead of 
$t$. Similarly, the expansion rate is given by Eq.(3) with the same 
change of the argument. It is interesting that this new expression for the
expansion rate is the same as for the hyperbolical trajectories considered above. 
Therefore the conclusions we made above extend directly to this new case. 

Our analysis of the trajectories and conclusions about the expansion
rate for the Local Volume are completely confirmed by computer models 
that trace back the observed kinematics of real galaxies of the local 
expansion flow (Karachentsev et al. 2003).

Summing up, we may say that an answer to the question in the title of 
the paper may be like this: the rate of the cosmological expansion in the 
Local Volume is similar to the global rate because the cosmic vacuum 
with its perfectly uniform density dominates the present-day dynamics
of the Hubble flow both locally and globaly .
The bulk motion does not affect this result basically because vacuum is
co-moving with any motion (see, for instance, Chernin 2001, Chernin et al. 2003).
        
This work was partially supported by
RFBR grant 01--02--16001.

{}


\newpage
\begin{figure}
\caption {Distribution in equatorial coordinates of 5272 galaxies with
radial velocities less than 2300 km/s. The Zone of Avoidance in the
Milky Way is shaded. The Virgo cluster (``V'') is situated near the map center.}
\end{figure}

\begin{figure}
\caption {Distribution in Supergalactic coordinates of the same 5272 galaxies.}
\end{figure}

\begin{figure}
\caption{Distribution in Cartesian supergalactic coordinates of nearby
galaxies with $V_{LG} <$ 1500 km/s. Central pannel: galaxies situated
within the Local supercluster plane; bottom and upper pannels: galaxies below
and above the Local supercluster plane, respectively.}
\end{figure}

\begin{figure}
\caption{Distribution of the Local volume galaxies within 6 Mpc around the
Milky Way, projected onto the the Supergalactic plane. The brightest
members of eight nearest groups are shown as asterisks.}
\end{figure}

\begin{figure}
\caption{Radial velocity - distance relation for 156 Local Volume galaxies.
The galaxies with acurate distance estimates are shown as filled circles,
and galaxies with less reliable distances are indicated as crosses. The
members of M 81 and Cen A groups are shown by open circles and squares.}
\end{figure}

\begin{figure}
\caption{The Hubble diagram for centroids of the eight nearest groups.}
\end{figure}

\begin{figure}
\caption{Different apex positions from Table 1 in galactic coordinates
(crosses). Positions of the Virgo attractor, the Great Attractor, the Shapley
concentration, as well as centroids of IRAS and 2MASS sources are shown as
grey circles.}
\end{figure}


\begin{thebibliography}{}

\bibitem{} Baryshev, Yu., Teerikorpi, P. 2002 {\em Discovery of Cosmic Fractals},(World
      Scientific Publishing Co.)

\bibitem{}Baryshev, Yu., Chernin A., Teerikorpi P., 2001, A\&A {\bf 378}, 729

\bibitem{}Busswell G.S., Shanks T., Outram P.J., et al. 2003,astro-ph/0302330

\bibitem{}Chernin A. 2001, Physics-Uspekhi, {\bf 44},1009

\bibitem{}Chernin A., Teerikorpi P., Baryshev Yu. 2003, Adv. Space Res. {\bf 31}, 459

\bibitem{}Dekel A., Eldar A., Kollat T. et al. 1999, ApJ, {\bf 522}, 1

\bibitem{}Dolgachev V.P., Domozhilova L.M., Chernin A.D. 2003, in press

\bibitem{}Ekholm T., Teerikorpi P., Theureau G. et al., 1999, A\&A {\bf 347}, 99

\bibitem{}Ekholm T., Baryshev Yu., Teerikorpi P., Hanski M., Paturel G. 2001,
      A\&A {\bf 368}, L17

\bibitem{}Freedman W.L., Madore B., Gibson B.K. et al., 2001, ApJ {\bf 553}, 47

\bibitem{}Giovanelli R., Haynes M.P., Freudling W., et al. 1998, ApJ {\bf 505}, L91

\bibitem{}Giovanelli R., Dale D., Haynes M., Hardy E., Campusano L. 1999,
     ApJ, {\bf 525}, 25

\bibitem{}Karachentseva V.E., Karachentsev I.D., 1998, A\&AS {\bf 127}, 409

\bibitem{}Karachentseva V.E., Karachentsev I.D., 2000, A\&AS {\bf 146}, 359

\bibitem{}Karachentsev I.D., Makarov D.I.,Sharina M.E. et al. 2003, A\&A, {\bf 398}, 493

\bibitem{}Karachentsev I., Sharina M, Makarov D. et al, 2002, A\&A {\bf 389}, 812

\bibitem{}Karachentsev I.D., Makarov D.I., 2001, Astrofizika {\bf 44}, 5

\bibitem{}Karachentsev I.D., Karachentseva V.E., Kudrya Yu.N.,
     Parnovsky S.L., 2000, Astron. Reports, {\bf 44}, 175

\bibitem{}Karachentsev I.D., Drozdovsky I., Kajsin S., et al. 1997, A\&AS, {\bf 124}, 559

\bibitem{}Karachentsev I., Makarov D., 1996, AJ {\bf 111}, 535

\bibitem{}Karachentsev I.D., Tikhonov N.A., 1994, A\&A, {\bf 286}, 718

\bibitem{}Karachentsev I.D., Karachentseva V.E., Parnovsky S.L., 1993,
      Astron. Nachr. {\bf 314}, 97 (FGC)

\bibitem{}Karachentsev I.D., Chernin A.D., Valtonen M.J. et al. 2003 (in press)

\bibitem{}Kraan-Korteweg R., Tammann G.A., 1979, Astron. Nachr. {\bf 300}, 181

\bibitem{}Kraan-Korteweg R.C., Lahav O., 2000, A\&ARv {\bf 10}, 211

\bibitem{}Kogut et al. 1993, ApJ {\bf 419}, 1

\bibitem{}Kudrya Yu.N., Karachentseva V.E., Karachentsev I.D. et al. 2003,
    A\&A, submitted
\bibitem{}Maller A.H., McIntosh D.H., Katz N., Weinberg M.D., 2003,
     astro-ph/0303592

\bibitem{}Paturel G., Bottinelli L., Di Nella H. et al. 1996, Catalogue of
 Principal Galaxies, Saint-Genis Laval, Observatoire de Lyon (LEDA)

\bibitem{}Peebles, P.J.E. 1993 {\em Principles of Cosmology} (Princeton: Princeton
    Univ. Press)

\bibitem{}Perlmuter S., Aldering G., Goldhaber G. et al. 1999, ApJ, {\bf 517}, 565

\bibitem{}Riess A.G., Filippenko A.V., Challis P. et al. 1998, AJ, {\bf 116}, 1009

\bibitem{}Rowan-Robinson M. et al. 2000, MNRAS, {\bf 314}, 375

\bibitem{}Ryan-Weber E., Koribalski B.S., Staveley-Smith L., et al. 2002,
	 AJ, 124, 1954

\bibitem{}Sandage, A. 1986, ApJ, {\bf 307}, 1

\bibitem{}Sandage, A. 1999, ApJ, {\bf 527}, 479

\bibitem{}Sandage A., Tammann G.A., Hardy E., 1972, ApJ {\bf 172}, 253

\bibitem{}Spergel D.N., Verde L., Peiris H.V., et al. 2003,astro-ph/0302209

\bibitem{}Sylos Labini, F., Montuori, M., Pietronero, L. 1998,
      Phys.Rep., {\bf 293}, 61

\bibitem{}Teerikorpi, P. 1997 Ann.Rev.Astron.Astrophys. {\bf 35}, 101

\bibitem{}Teerikorpi, P., Hanski, M., Theureau G. et al.
      1998, A\&A, {\bf 334}, 395

\bibitem{}Teerikorpi, P., Paturel, G. 2002 A\&A {\bf 381}, L37

\bibitem{}Theureau G, Hanski M, Ekholm T et al. 1997 A\&A {\bf 322}, 730

\bibitem{}Tikhonov A., Makarov D., Kopylov A. 2000,
      Bull. Special Astrophys. Obs., {\bf 50}, 39

\bibitem{}Tonry J.L. et al. 2000, ApJ, {\bf 530}, 625

\bibitem{}Tully B., 1988, Nearby Galaxy Catalog, Cambridge University Press

\bibitem{}Vaucouleurs G. de, Vaucouleurs A. de, Buta R.G. et al. 1991, Third
     Reference Catalog of Bright Galaxies, N.Y., Springer-Verlag

\bibitem{}Wu, K.K.S., Lahav, O., Rees, M.J. 1999, Nature {\bf 397}, 225

\end{thebibliography}
\end{document}